\documentclass[longauth]{aa}

\usepackage{txfonts}
\usepackage{graphicx}
\usepackage{natbib}

\begin{document} 

\title{The ALPINE-ALMA [CII] survey}
\subtitle{Survey strategy, observations,  and sample properties of 118 star-forming galaxies at 4<z<6}
          %\thanks{Based on data obtained with the ALMA observatory, under Large
          %Program 2017.1.00428.L }

\author{O.~Le F\`evre\inst{1}
% ALPINE co-PIs
\and M. B\'ethermin \inst{1}
\and A.~Faisst\inst{2}
\and G.C.~Jones \inst{15,16}
\and P.~Capak\inst{2,27,28}
\and P.~Cassata\inst{4,5}
\and J.D.~Silverman\inst{6,32}
\and D. Schaerer\inst{7,8}
\and L. Yan\inst{23}
% ALPINE co-Is:
\and R.~Amorin\inst{12,13}
\and S.~Bardelli\inst{11}
\and M. Boquien\inst{14}
\and A.~Cimatti\inst{9,10}
\and M.~Dessauges-Zavadsky \inst{7}
%\and J. ~Dunlop\inst{14}
\and M.~Giavalisco\inst{17}
\and N.P.~Hathi\inst{17}
\and Y. Fudamoto \inst{7,25}
\and S. Fujimoto \inst{25,26}
\and M. Ginolfi \inst{7}
\and C. Gruppioni\inst{11}
\and S.  Hemmati \inst{28}
%\and Hughes \inst{1}
\and E. Ibar \inst{20}
\and A.~Koekemoer\inst{17}
\and Y.~Khusanova\inst{1}
\and G.~Lagache\inst{1}
\and B.C.~Lemaux\inst{3}
\and F. Loiacono\inst{9,11}
\and R.~Maiolino \inst{15,16}
\and C. ~Mancini\inst{4,5}
\and D. ~Narayanan\inst{27,30,31}
\and L. Morselli\inst{4}
\and Hugo M\'endez-Hern\`andez\inst{20}
%\and D.~Masters \inst{1}
%\and Nagao \inst{1}
%\and Narayanan \inst{1}
\and P.A.~Oesch \inst{7}
%\and Pavesi \inst{1}
\and F. ~Pozzi \inst{9}
\and M. Romano\inst{4}
%\and B.~Ribeiro \inst{1}
\and D. Riechers\inst{29,22}
%\and Rujopakarn\inst{1}
\and N. Scoville\inst{33}
\and M.~Talia\inst{9,11}
\and L. A. M.~Tasca\inst{1}
\and R.~Thomas\inst{18}
\and S.~Toft\inst{27,28}
\and L.~Vallini\inst{21}
\and D.~Vergani\inst{11}
\and F.~Walter\inst{23}
%\and Wei-Hao Wang\inst{1}
\and G.~Zamorani \inst{11}
\and E.~Zucca\inst{11}
}

%1
\institute{Aix Marseille Universit\'e, CNRS,  CNES, LAM (Laboratoire d'Astrophysique de Marseille), 13013, Marseille, France \email{olivier.lefevre@lam.fr and matthieu.bethermin@lam.fr}
\and
%2
IPAC, California Institute of Technology, 1200 East California Boulevard, Pasadena, CA 91125, USA
%3
\and
Department  of  Physics,  University  of  California,  Davis,  One  Shields  Ave.,  Davis,  CA  95616,  USA
\and
%4
Dipartimento di Fisica e Astronomia, Universit\`a di Padova, Vicolo dell’Osservatorio, 3 35122 Padova, Italy
\and
%5
INAF, Osservatorio Astronomico di Padova, vicolo dell’Osservatorio 5, I-35122 Padova, Italy
\and
%6
Kavli Institute for the Physics and Mathematics of the Universe, The University of Tokyo, Kashiwa, Japan 277-8583 (Kavli IPMU, WPI)
\and
%7
Department of Astronomy, University of Geneva, ch. des Maillettes 51, CH-1290 Versoix, Switzerland
\and
%8
Institut de Recherche en Astrophysique et Plan\'etologie - IRAP, CNRS, Universit\'e de Toulouse, UPS-OMP, 14, avenue E. Belin, F31400 Toulouse, France
\and
%9
University of Bologna, Department of Physics and Astronomy (DIFA), Via Gobetti 93/2, I-40129, Bologna, Italy
\and
%10
INAF - Osservatorio Astrofisico di Arcetri, Largo E. Fermi 5, I-50125, Firenze, Italy
\and
%11
INAF - Osservatorio di Astrofisica e Scienza dello Spazio di Bologna, via Gobetti 93/3, I-40129, Bologna, Italy
\and
%12
Instituto de Investigacion Multidisciplinar en Ciencia y Tecnologia, Universidad de La Serena, Raul Bitran 1305, La Serena, Chile
\and
%13
Departamento  de Astronomia,  Universidad  de  La
Serena, Av.  Juan Cisternas 1200 Norte, La Serena, Chile
\and
%14
Centro de Astronomia (CITEVA), Universidad de Antofagasta, Avenida Angamos 601, Antofagasta, Chile
\and 
%15
Cavendish Laboratory, University of Cambridge, 19 J. J. Thomson Ave., Cambridge CB3 0HE, UK
\and
%16
Kavli Institute for Cosmology, University of Cambridge, Madingley Road, Cambridge CB3 0HA, UK
\and
%17
Space Telescope Science Institute, 3700 San Martin Drive, Baltimore, MD 21218, USA
\and
%18
European Southern Observatory, Av. Alonso de C\'ordova 3107, Vitacura, Santiago, Chile 
\and
%19
Astronomy Department, University of Massachusetts, Amherst, MA 01003, USA
\and
%20
Instituto de F\'isica y Astronom\'ia, Universidad de Valpara\'iso, Avda. Gran Breta\~na 1111, Valpara\'iso, Chile
\and 
%21
Leiden Observatory, Leiden University, PO Box 9500, 2300 RA Leiden, The Netherlands
\and
%22
Max-Planck Institut f\"ur Astronomie, K\"onigstuhl 17, D-69117, Heidelberg, Germany
%23
\and
Department of Astronomy, California Institute of Technology, 1200 E. California Blvd., MC 249-17, Pasadena, CA 91125, USA
\and
%24
Waseda University, Department of Physics, Waseda Research Institute for Science and Engineering Tokyo, Japan
\and
%25
Research Institute for Science and Engineering, Waseda University, 3-4-1 Okubo, Shinjuku, Tokyo 169-8555, Japan
\and
%26
National Astronomical Observatory of Japan, 2-21-1, Osawa, Mitaka, Tokyo, Japan
\and
%27
Cosmic Dawn Center (DAWN), Copenhagen, Denmark
\and
%28
Niels Bohr Institute, University of Copenhagen, Lyngbyvej 2, DK-2100 Copenhagen, Denmark
\and 
%29
Department of Astronomy, Cornell University, Space Sciences Building, Ithaca, NY 14853, USA 
\and 
%30
Department of Astronomy, University of Florida, 211 Bryant Space Sciences Center, Gainesville, FL 32611 USA
\and
%31
University of Florida Informatics Institute, 432 Newell Drive, CISE Bldg E251, Gainesville, FL 32611
\and 
%32
        Department of Astronomy, School of Science, The University of Tokyo, 7-3-1 Hongo, Bunkyo, Tokyo 113-0033, Japan
\and 
%33
Cahill Center for Astrophysics, California Institute of Technology, 1216 East California Boulevard, Pasadena, CA 91125, USA}

\date{Received ; accepted }

% \abstract{}{}{}{}{} 
% 5 {} token are mandatory
 
  \abstract{The ALMA-ALPINE [CII] survey is aimed at characterizing the properties of a sample of normal star-forming galaxies (SFGs).  The ALMA Large Program to INvestigate (ALPINE) features 118 galaxies observed in the [CII]-158$\mu$m line and far infrared (FIR) continuum emission during the period of rapid mass assembly, right after the end of the HI reionization, at redshifts of 4<z<6. We present the survey science goals, the observational strategy, and the sample selection of the 118 galaxies observed with ALMA, with an average beam minor axis of about 0.85\arcsec, or $\sim$5 kpc at the median redshift of the survey.
 The  properties of the sample are described, including spectroscopic redshifts derived from the UV-rest frame,  stellar masses, and star-formation rates obtained from a spectral energy distribution (SED) fitting. The observed properties derived from the ALMA data are presented and discussed in terms of the overall detection rate in [CII] and FIR continuum, with the observed signal-to-noise distribution. The sample is representative of the SFG population in the main sequence at these redshifts. The overall detection rate in [CII] is 64\% for a signal-to-noise ratio (S/N) threshold larger than 3.5 corresponding to a 95\,\% purity (40\,\% detection rate for S/N$>$5). Based on a visual inspection of the [CII] data cubes together with the large wealth of ancillary data, we find a surprisingly wide range of galaxy types, including 40\% that are mergers, 20\% extended and dispersion-dominated, 13\% compact, and 11\% rotating discs, with the remaining 16\% too faint to be classified. This diversity indicates that a wide array of physical processes must be at work at this epoch, first and foremost, those of galaxy mergers. This paper sets a reference sample for the gas distribution in normal SFGs at 4<z<6, a key epoch in galaxy assembly, which is ideally suited for  studies with  future facilities, such as the James Webb Space Telescope (JWST) and the Extremely Large Telescopes (ELTs). }


\authorrunning{Le F\`evre, O., and ALPINE team }

\titlerunning{The ALPINE-ALMA [CII] survey: presentation}

\maketitle

\section{Introduction}
\label{intro}

The mass assembly in galaxies at different epochs is the result of several physical processes which, in combination, produce the remarkable observed evolution of the  star formation rate density (SFRD) with cosmic time \citep[][and references therein]{Silk2012,madau:14,dayal2018}. The SFRD first rises during the reionization epoch to reach its peak at z$\sim2-3$ following a $\sim1$ dex increase in $\sim3$ Gyr. It then decreases by $\sim 0.8$ dex in $\sim10$ Gyr up to the current point in time \citep{madau:14,Bouwens:14}. Along with star formation, the total stellar mass density (SMD) in galaxies is observed to rise steeply from early timescales to z$\sim$2, followed by a slower increase at z$<$2 \citep{Ilbert:13}.

The key element at the root of the SFRD and SMD evolution is the transformation of gas into stars within a hierarchical picture of galaxy assembly. Two main processes are shown, thanks to increasingly detailed simulations, to drive this evolution: gas accretion and galaxy merging \citep{Hopkins2006,Dekel:09,Bournaud:11,Naab2017}. We expect this to be  tempered by feedback processes from gas expelled from galaxies by strong AGN or stellar jets and winds \citep{Silk:97,Hopkins2008,silk2013}. While this is appealing from the standpoint of theory and simulation, there is actually very little observational evidence for  a comprehensive, consistent, and quantitative picture, particularly among the early cosmic epochs, when the major phase of mass assembly is underway. Galaxy mergers, major and minor, are observed at all epochs \citep[e.g.,][]{Conselice2014}, with a major merger rate increasing from the local universe to $z\sim2$ \citep[e.g.,][]{Lotz2011,Lopez-Sanjuan:13,Mantha2018}, and possibly flattening to $z\sim4-5$ \citep[e.g.,][]{Tasca:14,ventou2017}, while gas accretion suffers from weak signatures that are difficult to identify observationally, and its effects are only identified indirectly \citep[e.g.,][]{Bouche:13}. On the other hand, feedback processes are directly measured \citep[e.g.,][]{LeFevre2019} and thought to affect both the bright and faint  end of the galaxy luminosity function \citep[LF; e.g.,][]{Croton:2006,Hopkins2008,Gabor:2010,Gabor:2011}, which is due primarily to AGN and stellar processes, respectively.

To disentangle the relative contributions of these processes, the far infrared (FIR) domain that is redshifted in the sub-mm for high-z galaxies  is proving to be a particularly rich resource of information. From the sub-mm, it is now possible to investigate the properties of star-forming galaxies up to the epoch of HI reionization. The [CII]-158$\mu$m line is the dominant coolant, making it one the strongest FIR lines. The  [CII] emission is primarily coming from photo-dissociation regions (PDR) and cold neutral medium (CNM) of molecular clouds. [CII] at high-z has raised considerable interest as it probes the gas which stars form out of in normal galaxies \citep[e.g.,][and references therein]{Ferrara2019}. It also broadly traces star formation activity, offering an important window on galaxy formation \citep{Carilli2013, DeLooze2014}. This has led to the detection of strong [CII] emitters, up to very high redshifts \citep{Capak2015,Carniani2017}, which is an easier measurement than the FIR continuum.
Searching for [CII] emission, interpreting and simulating the observations, and comparing them with other  emission lines, such as  Lyman-$\alpha$, has therefore become a major new way of studying high-z galaxies. The strong UV radiation in high-z galaxies results in a non-negligible fraction of [CII] emission from the extended warm interstellar medium \citep[ISM; e.g.,][]{Capak2015, Faisst2017}. The  [CII] emission together with its morphology provides important information on the SFR and ISM properties \citep[e.g.,][]{Vallini2015,Wellons2016, Olsen2017}. The FIR continuum emission adjacent to [CII] is near the peak of the FIR emission. It constrains the total FIR luminosity and provides a  measurement of the total SFR when combined with UV continuum measurements. It can also be combined with UV colors and luminosity to construct the infrared-excess (IRX, defined as $L_{FIR} /L_{UV}$) versus UV color  diagnostic, providing insights into the spatial distribution of dust, dust grain properties, and metallicity \citep{Reddy2012,Faisst2017}.

Simulations of  galaxy formation during and right after reionization provide information on the possible properties of these galaxies despite the difficulty in taking into account early galaxy-formation processes during and right after the Epoch of reionization  in a consistent way \citep[e.g.,][]{Dayal2013,Maiolino2015}. Specific predictions related to [CII] emission are useful for guiding and making comparisons  with the observations \citep[e.g.,][]{Yue2015,Vallini2015,Olsen2017,Kohandel2019}.

This whole domain opened up at high redshift (z>4) with the ALMA interferometer becoming fully operational when pilot observations took note that detecting [CII] for normal galaxies was ubiquitous even with short on-source exposure times \citep{Capak2015}. Galaxies with star formation rates as low as a few $M_{\odot}$\,yr$^{-1}$ have been detected in [CII] at z$\sim5$ \citep{Riechers2014,Capak2015}, and [CII] is now detected for galaxies well into the reionization epoch  \citep[e.g.,][]{bradac2017,Smit2018,Harikane2018,Fujimoto2019, sobral2019,Hashimoto2019}.

However, existing observations of [CII] in normal galaxies at these epochs are still scarce. As strong sub-mm sources have primarily been  targeted \citep[e.g.,][]{Maiolino2009,Carilli2013,Wagg2012,Riechers2014}, they provide us with a view biased towards the intensely star-forming population with SFR$>1000$ $M_{\odot}$\,yr$^{-1}$. Normal galaxies, that is, galaxies with SFR in the range from $\sim 10$ up to a few hundred solar masses per year, lying on the so-called main sequence at these redshifts \citep[e.g.,][]{Speagle2014,tasca2015,Tomczak2016,Pearson2018,Khusanova2019}, have not been observed in statistically representative numbers. The \cite{Capak2015} observations proved that this was feasible and prompted us to submit the ALPINE Large Program, which has been largely designed based on the properties of the \cite{Capak2015} sample. A key element was the availability of large samples of these normal galaxies, with accurate spectroscopic redshifts{\it } \citep{LeFevre:15,Hasinger2018}, to be capable of defining ALMA observations with a high success rate in detecting [CII].

This paper presents a general overview of  the survey, in combination with papers presenting a detailed account of the data processing, [CII] flux and continuum measurements \citep{Bethermin:20}, and ancillary data with physical parameters computation \citet{Faisst:19}. The layout of this paper is as follows. In Sect. \ref{design}, we present the general ALPINE survey design, with  our science goals and sample selection as in the original proposal. We give an overview of  the ALMA observations, as well as a summary of the large amount of ancillary observations  in Sect.\ref{obs}. In Sect.\ref{sample}, we describe the main properties of the sample, including the redshift distribution, detection rates in [CII] and continuum, and observed flux limits. Flux maps of all sample galaxies in the [CII] line are presented in Sect.\ref{c2maps}. We use these maps, kinematic data, and all ancillary imaging data   to perform an empirical visually-based morpho-kinematic classification as described in Sect. \ref{class}. We summarize our findings in Sect.\ref{summary}.
Throughout the paper, we use a $\Lambda$CDM cosmology with $H_0=70$ km/s/Mpc, $\Omega_{\Lambda}=0.70$, $\Omega_m=0.30$. All magnitudes are given in the AB system.

%--------------------------------------------------------------------

\section{ALPINE survey design}
\label{design}

\subsection{Science goals}

The main  goals of ALPINE at 4<z<6 are broadly defined as follows:

\textbf{Characterize the use of [CII] as a SFR indicator at these epochs.} The prevalence of [CII] in high-z galaxies is a promising tool to estimate SFRs of FIR continuum-faint galaxies. While local studies find a good correlation between [CII] and SFR, this relation may change at the lower metallicities of high-z galaxies. There is also evidence that [CII] is often emitted from the diffuse CNM or HII regions in addition to PDRs \citep{Herrera2017,Pineda2013,Vallini2015}, and it also traces the difuse ionized gas; see \citep{Pavesi2016,Pavesi2019}. ALPINE allows calibrating this relation by comparing [CII] derived SFR to other indicators (FIR, UV, SED) over a large range of physical properties.

\textbf{A comprehensive and precise (accuracy better than 20\%) measurement of the SFRD at these epochs from the UV+FIR continuum and [CII] emission} allows us  to constrain the mechanisms which fuel the initial growth of typical galaxies in the early universe. The total SFRD at  z>4, a key epoch in  galaxy assembly, is a crucial element in understanding galaxy formation. However, it remains a difficult observational measurement, as we do not yet know how much of the star formation is hidden from the wealth of existing deep UV observations. Only a survey of selected sources based on the FIR emission could solve the question but this is currently not feasible for a statistically representative sample given the small field of view of the ALMA telescope. With ALPINE, we follow a stepped approach. Starting from well-studied sources in the UV, the goal is to obtain FIR continuum and [CII] measurements to measure the fraction of their star formation that is hidden by dust. Combined with the SFR derived from the UV continuum, this would deliver the total star formation of UV-selected samples.

\textbf{Estimate the remaining fraction of the star formation that may not traced by UV sign-posts.} In addition to the above, we aim to use the serendipitous survey assembled by ALPINE on a total area of about 25 square arcminutes to estimate what fraction of star formation is missed in obscured sources. Together these measurements will result in a first estimate of the [CII] luminosity function and the total SFRD at 4<z<6 consolidated from the UV and FIR. 

\textbf{The first detailed characterization of the ISM properties using $L_{FIR} / L_{UV}$ and [CII] /FIR diagnostics.} The evolution of [CII] emission and its resolved velocity profile provide important information on the SFR and ISM properties, setting constraints
on the dynamical and gas masses of galaxies. The morphology of the [CII] emission, and more generally of the gas distribution, indicates whether star formation is
compact or extended, which is an important element to understand with regard to high-z star formation \citep[e.g.,][]{Wellons2016}.
Beyond the line flux, the FIR continuum emission constrains  the total FIR luminosity and provides a good measurement of the total SFR when combined with UV continuum measurements. The continuum flux can also be combined with UV color and luminosity to
construct the IRX-$\beta$ diagnostic, providing insight into the spatial
distribution of dust, dust grain properties, and metallicity \citep{Reddy2012, Faisst2017}. The [CII] line has also been used as a tracer of the molecular gas content  \citep{Hughes2017,Zanella2018}, which is shown to be more reliable in low-metallicity environments than, for example, CO.

\textbf{Star-forming main-sequence and merger rates at z > 4.} The SFRD across cosmic time as well as
the shape and scatter of the star forming main sequence (MS) in the SFR-stellar mass plane provide important constraints on the starburst duty cycle and
merger rate of galaxies \citep{Rodighiero2011,Guo13,Tomczak2016,Tacchella2015} and ultimately their mass growth \citep[cold accretion vs. merger paradigm][]{Dekel:09,Dave:11,Tasca:14,Faisst2017}. The independent SFR measurements from FIR and [CII] in
addition to  the SED fitting is meant to help constrain the true scatter of the MS at z>4 as a function of stellar mass. The comparison to H$\alpha$ based SFRs on <100 Myr timescales from Spitzer colors \citep{Faisst2016}, or later from JWST spectroscopy, is aimed at allowing us to put constraints on the starburst duty cycles.
Following the  serendipitous [CII] detections of UV-faint galaxies in interaction with the main targets
\citep[extrapolating from][we expected to find $\sim$15-30 mergers in ALPINE]{Capak2015}, which allow us to model the merger rates and thus
constrain the dominant mode of mass build-up of galaxies in the early universe.

\textbf{A first measurement of dynamical masses from spectrally resolved  [CII]}, combined with stellar
masses and statistical estimates of dark matter halo masses to measure dust content, gas fraction,
and their evolution.
The [CII] line is an excellent tracer of gas dynamics, which can be used in
the same way as H$\alpha$ at z<2 \citep[e.g.,][]{Forster2009,Epinat2012,Molina2017}. The ALPINE observations are aimed at providing a first
approach at dynamical measurements at 4<z<6, deep into the epoch of early galaxy formation.
Dynamical masses (M$_{dyn}$ ) can be estimated from the velocity dispersion, $\sigma_{vel}$ , derived from the [CII] line emission, while the brightest galaxies resolved in
[CII] enable a direct measurement of the [CII] extent. It is expected that resolved galaxy sizes \citep[][]{Ribeiro2016} lead to a useful dynamical mass measurement for $\sim20$\% of the sample. Upper limits are derived for the rest of the sample. 

\textbf{Gas fractions at z>4:} Comparing $M_{dyn}$ to reliable emission-line-corrected stellar masses from deep
Spitzer imaging \citep{Laigle2016,Faisst:19} and gas mass estimates from FIR measurements \citep[e.g.,][]{Scoville2016} puts the first direct constraints on the gas fractions f$_{gas}$ at z>4. The contribution of dark matter to the
total velocity component at 1-2\,$R_e$, where $R_e$ is the effective radius of the galaxy, is expected to be low \citep{Barnabe2012}, but can be estimated by
matching our sample to the output of state-of-the-art hydro-dynamical simulations such as EAGLE \citep{McAlpine2016} FIRE \citep{Hopkins2014}, or Illustris TNG \citep{Dave2019}, or estimating average dark halo masses from HOD modeling of the correlation function \citep{Durkalec15}. Combined with SFRs
derived from UV, FIR, [CII], and H$\alpha$, this will constrain SFR efficiencies and gas depletion times, thus
providing insight into the emergence and growth of massive galaxies \citep{Tacconi2013,Genzel2015,Dessauges2017,Tacconi2018}.

\textbf{The role of feedback processes in the early Universe.}
The prevalence of feedback processes from galactic winds produced by massive stars, supernovae and AGN are studied using kinematic diagnostics and [CII] line profiles \citep[see from ALPINE, ][]{Ginolfi2019}.

\subsection{Sample selection}

The sample is drawn from large spectroscopic survey samples of normal SFGs in the COSMOS \citep[][]{Scoville:07a} and ECDFS fields \citep[][]{Giacconi2002}. 

\begin{figure}
   \centering
  \includegraphics[width=\hsize]{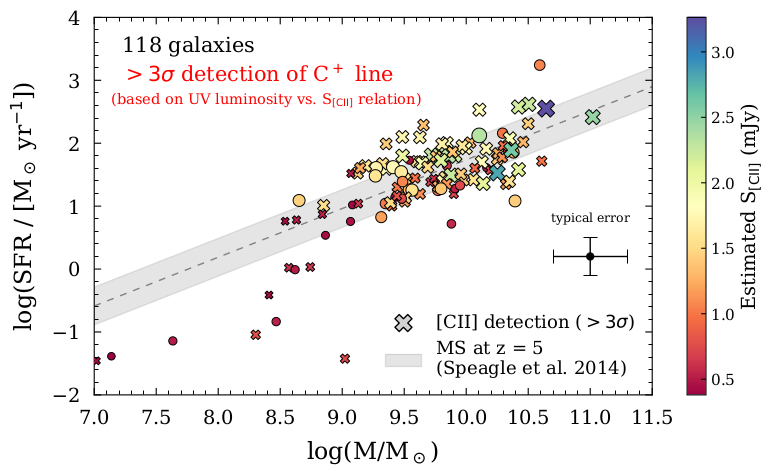}
      \caption{Stellar mass vs. SFR distribution of ALPINE sources as selected prior to ALMA observations, colored by expected [CII] line peak flux estimated
 from \citet{Capak2015}. M$_{star}$ and SFR were obtained from SED fitting of the multi-wavelength photometry available at the start of this program (see \citealt{Faisst:19} for more details). The cross represents the average
1$\sigma$ uncertainty. The large proposed sample aims to
quantify average trends over a large range of galaxy properties.  Those galaxies detected at >3$\sigma$ in [CII] are identified with crosses. ALPINE provides [CII] emission line
measurements at >3.5$\sigma$ for 63\% of these galaxies (see text). 
              }
         \label{Fig_mass_sfr}
   \end{figure}

A key element of the selection is based on galaxies having a reliable spectroscopic redshift  in $4.4<z_{spec} <5.9$ ($<z_{spec} > \sim4.7$), excluding $4.65<z<5.05$ (where [CII] falls in a low transmission atmospheric window). Galaxies are UV-selected \citep[see][for more details]{LeFevre:15} with $L_{UV} >0.6L^*$ to include most of the star formation traced by the UV, and excluding type I AGN identified from broad spectral lines. Accurate redshifts come from extensive spectroscopic campaigns at the VLT \citep[VUDS, ][]{LeFevre:15} and Keck \citep[DEIMOS, ][]{Hasinger2018}. While the VUDS sample is unbiased against Lyman-$\alpha$ emitters or absorbers \citep{Khusanova2019}, Lyman-$\alpha$ emitters may be over-represented in the DEIMOS sample; this will need to be taken into account in subsequent analysis of the whole ALPINE sample. Galaxies in the parent sample were mostly selected based on their photometric redshifts based on SED fitting, followed by UV spectroscopy to secure the redshift (see \citealt{LeFevre:15,Hasinger2018} for more details). The absolute UV luminosity cut (M$_{UV} <-20.2$) is equivalent to SFR>10 M$_{Sun}.yr^{-1}$, as seen in Fig.~~\ref{Fig_mass_sfr}. Assuming the \cite{DeLooze2014} relation, this SFR limit is equivalent to $L_{[CII]} > 1.2\times10^8 L_{\odot}$. Some galaxies with SFR below $\sim 1$ solar masses per year were included when made possible by the observational setup. This sample is representative of the overall SFG population, rather than  ultra luminous infrared galaxies (ULIRGS), that is, mostly of galaxies  positioned on or near the so-called main sequence in the SFR versus M$_{star}$ plane observed at these redshifts \citep[e.g.,][]{Speagle2014,tasca2015,Tomczak2016,Pearson2018,Khusanova2019}, with M$_{star}$ and SFR derived from SED fitting setting the redshift to the spectroscopic redshift. We selected 118 galaxies based on these criteria. More details on the sample properties are given in Sect.~\ref{sample}.

%----------------------------------------------------------------- 
%-----------------------------------------------------------------

\section{Observational data}
\label{obs}

\subsection{ALMA}

Here we summarize the observations, pointing to \citet{Bethermin:20} for an extensive description of observations and data processing, following the best practice with customized pipelines based on  the  Common Astronomy Software Applications (CASA) tools \citep{McMullin2007}.

This program was awarded an ALMA Large Program status under number 2017.1.00428.L for a total of 69 hours, including on-source time, calibrations, and overheads. ALMA observations were carried out in Band-7 starting in May 2018 during Cycle 5 and  completed in February 2019 in Cycle 6. Each target was observed for about 30 minutes and up to one hour of integration time, with the phase centers pointed to the UV rest-frame positions of the sources. The availability of spectroscopic redshifts allowed us to accurately set the main spectral window on the expected [CII]  frequencies. In order to minimize overheads, targets making use of a similar setup were grouped. Two spectral windows were placed in the side band containing the [CII] lines of the two grouped sources. In addition, we placed two additional spectral windows to cover the other side band. They are used for FIR continuum measurements, an important component of this program as continuum measurements can be used as a proxy for the total FIR, therefore, for the SFR. The main calibration for the [CII]-SFR relation partly relies on those galaxies with continuum data.

 At these redshifts, the velocity width of one ALMA bandpass in band-7 is as narrow as $\sim$3000 km/s per sideband. Samples with photometric redshifts accurate to $\sim 0.05\times (1+z)$ at these redshifts \citep[e.g.,][]{Ilbert:13} would have added a considerable uncertainty on the detection of [CII] and the associated incompleteness, making it more hazardous to build, for example, the [CII] luminosity function. With an accuracy of a few hundred km/s even at low spectral resolution from optical (UV rest-frame) spectroscopy \citep{LeFevre:15},  the availability of $z_{spec}$ is therefore a key element of this program that would ensure a high [CII] detection rate  and, for those galaxies which would be undetected, setting stringent upper limits.

We use the TDM mode of the ALMA correlator, which offers the largest bandwidth to optimize the continuum sensitivity. The resolution varies with redshift from 26 to 35 km/s. We assume 235 km/s FWHM line width (or sigma $\sim100$ km/s), which is the average width measured in the \cite{Capak2015} sample. Emission lines were thus expected to be spectrally resolved, giving the possibility to measure the line width when the signal-to-noise ratio (S/N) is sufficient. 

We prioritized detection over spatial resolution, and with the typical size of the \cite{Capak2015} sources being 0.5-0.7 arcsec, we elected to use ALMA array configurations (C43-1 or C43-2), offering a beam size not smaller than 0.7 arcsec to prioritize detection over spatial resolution. The minimum achieved minor-axis size of the ALMA beam is 0.72\,arcsec and the average beam size is 0.85"$\times$1.13".

We provide the ALPINE source list in Table \ref{tab_prop}, including  (RA, DEC) positions, spectroscopic redshift, [CII] S/N, when detected above $3.5\sigma$. This is the threshold at which our simulation indicates a 95\% reliability. As explained in \citet[][see Sect. 6.2 for a full description of the purity estimate]{Bethermin:20}, where the data analysis process and the validation simulations are described, we can use this low detection threshold because we are searching for S/N peaks close both spatially and spectrally from candidates and the number of detectable objects is high. In other words, searching for a line
in a reduced number of beams and channels reduces the expected number of spurious detections from noise fluctuations.

\subsection{Ancillary data}
\label{ancillary}

The choice of the COSMOS and ECDFS fields for the selection of the sample is driven by the availability of $z_{spec}$ as described above and also by the large suite of multi-wavelength data available in these fields, as is extensively described in \citet{Faisst:19}. These data enable us to construct the SED for each galaxy, which is essential for measuring the fundamental physical parameters needed to assess the general properties of the observed galaxy population.

All ALPINE targets in the COSMOS field \citep{Scoville2007} are covered by HST F814W i-band imaging \citep{Koekemoer:07},
Subaru optical imaging \citep{Taniguchi2007}, deep NIR YJHK imaging from the UltraVista Survey \citep{McCracken:12}, SPITZER 3.6 and 4.5 $\mu$m imaging \citep{Sanders:2007}, with good X-ray coverage with both XMM-Newton \citep{Hasinger2007} and Chandra \citep{Elvis2009,Civano2016}, as well as radio waves \citep{smolcic2017}. While most of the imaging is done under  point-spread functions (PSF) with similar spatial resolution as the ALMA imaging, typically FWHM$\sim 0.8$ arcsec, the HST F814W i-band imaging provides a sharper look with a  PSF$\sim$0.1 arcsec. A source catalog with  matched photometry is available as described in \cite{Laigle2016}.
The data that we use in the area of the ECDFS is from the CANDELS survey \citep{Grogin:2011, Koekemoer:11}. The derivation of key physical quantities from the SED fitting, including stellar mass, SFR, dust extinction, is described in detail in \citet{Faisst:19}.

%--------------------------------------------------------------------
\section{Sample properties}
\label{sample}

\subsection{Redshift distribution}

\begin{figure}
   \centering
  \includegraphics[width=\hsize]{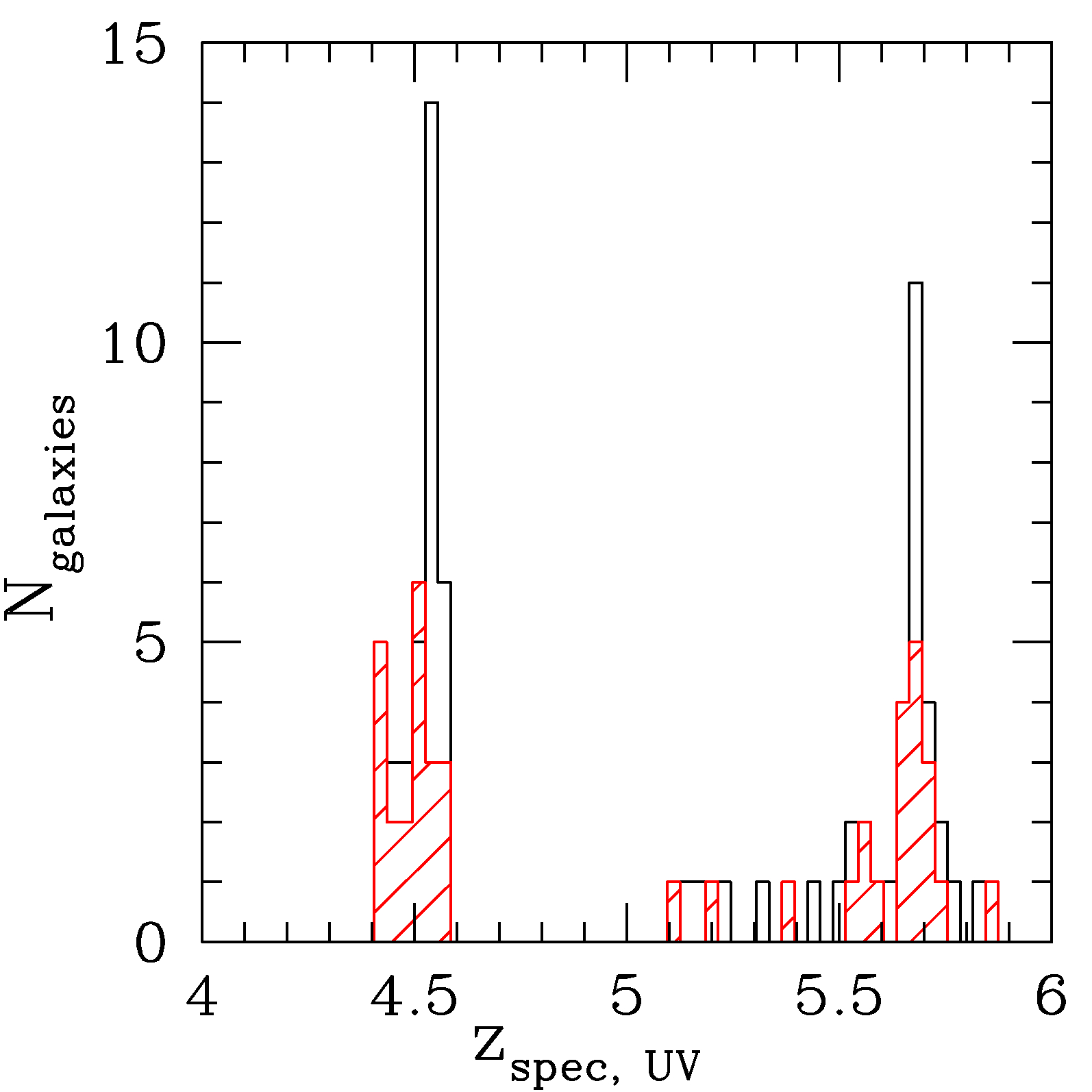}
      \caption{Redshift distribution of the ALPINE sample, using the spectroscopic redshifts measured from the UV rest-frame spectra  \citep{LeFevre:15,Hasinger2018}. The empty histogram is for all observed sources while the red shaded histogram is for those sources with [CII] measured at more than 3.5$\sigma$ above the noise.
              }
         \label{Fig_z_dist}
   \end{figure}

The redshift distribution of the ALPINE sample  is presented in Fig. \ref{Fig_z_dist} for all galaxies as well as those with [CII] detected at more than 3.5$\sigma$ above the noise. There are two separate peaks in the N(z) owing to the atmospheric visibility windows. 

Comparing the redshifts obtained from the UV spectra to those obtained from [CII], which defines the systemic redshift of the gas component, allows us to probe the velocity difference between the stellar and gas components. This is explored in \citet{Faisst:19} and \citet{Cassata:20}.

%----------------------------------------------------------------- 
%-----------------------------------------------------------------

\subsection{Detection rate in [CII] and continuum}

\begin{figure}
   \centering
  \includegraphics[width=\hsize]{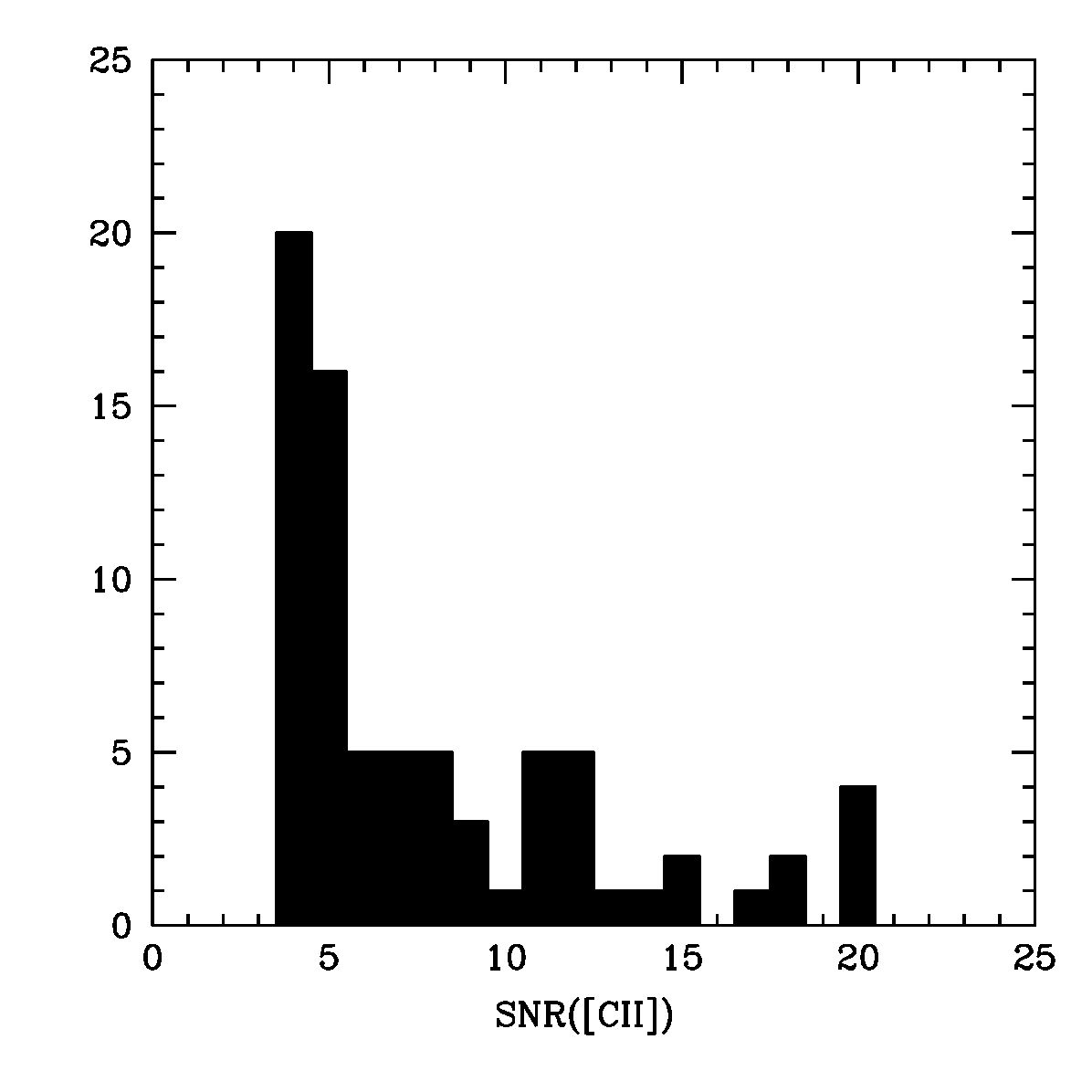}
      \caption{S/N of the integrated [CII] line flux for those sources with [CII] measured at more than 3.5$\sigma$ above the noise. 
              }
         \label{Fig_snr}
   \end{figure}

The S/N of the integrated [CII] line flux detections are presented in Fig.\ref{Fig_snr}. The median S/N is $\simeq6.2$ for the detected objects.  See \citet{Bethermin:20} for more details.
Taking 3.5$\sigma$ as a conservative detection limit (purity > 95\%) owing to the somewhat correlated noise of ALMA interferometric imaging, ALPINE detected [CII] in 75 galaxies out of 118, hence a success rate of 64\,\% (40\,\% for S/N$>$5) as presented in Table \ref{tab_prop}. In the continuum adjacent to [CII], 25 galaxies, or 21\% (9\,\% for S/N$>$5), are detected with S/N$>$3.5 (95\,\% purity, \citealt{Bethermin:20}). These rates are quite impressive given the redshift of the sources and short integration times. The lower continuum detection rate is as expected from the SED models in the FIR \citep{Bethermin2017}. Stacking of the continuum data will allow to place useful constraints on the fainter emitters. The S/N of most other targets varies in the range from 0.5 to 3, providing useful upper limits.

%The flux distribution of the [CII] line is presented  in  Fig. \ref{Fig_snr}. The median observed flux is $Flux([CII])\simeq0.71 Jy.km.s^{-1}$. See B\'ethermin et al. (2019, in prep.) for more details.

We identified a number of line emitters detected serendipitously in the data cubes. These are to be\ matched with the ancillary source catalogs in these fields in the aim of identifying whether the line is [CII] at the redshift of the targeted sample, or some other line at lower redshift (Loiacono et al., in prep.).

%----------------------------------------------------------------- 
%-----------------------------------------------------------------

\subsection{Positional offsets between [CII] and UV rest-frame}

Positional offsets between the [CII] flux distribution and the UV rest-frame images have been reported for galaxies at similar redshifts as the ALPINE sample \citep[e.g.,][]{Carniani2017}. Offsets from frame centers representing the UV position are clearly evident for some objects on Figs. ~\ref{Fig_class1} through \ref{Fig_class4}.
 A preliminary analysis shows that for most (90\%) of the sources the differences in RA and DEC are well represented by Gaussians centered at zero offset with a sigma of about 0.18", which is consistent with the typical uncertainty in position of the ALMA sources, and there is a significant offset (up to a maximum of $\sim1\arcsec$) present only for 10\% of the sources.
%A preliminary analysis shows a global offset with sigma(RA) = sigma(DEC)$\sim$0.18\arcsec. For most of the sources (~90\%) the data are consistent with no appreciable offset between the two components and a significant offset is present only for ~10\% of the sources. 
See \citet{Faisst:19} for more details.

%--------------------------------------------------------------------
\section{[CII] flux maps}
\label{c2maps}

\begin{figure*}
   \centering
  \includegraphics[width=\hsize]{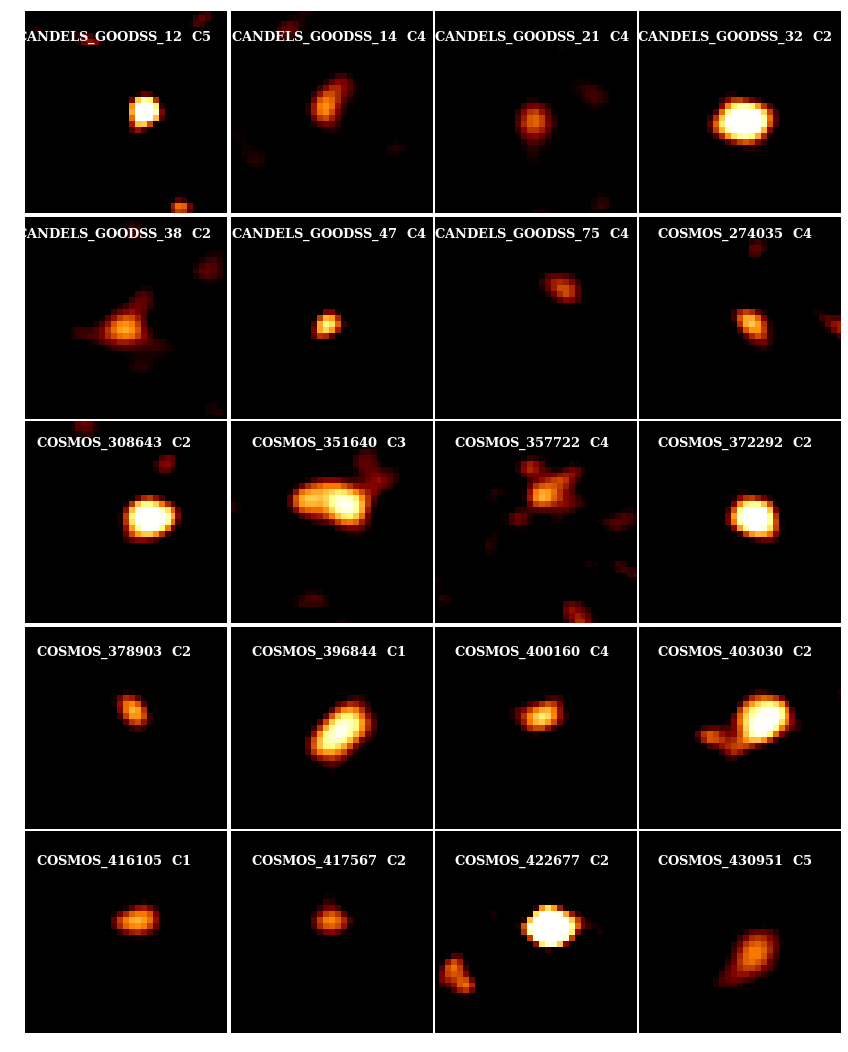}
      \caption{Velocity-integrated [C II] flux maps obtained collapsing the cube channels containing the [C II] line (see text). Each panel is $5\arcsec \times 5\arcsec$ or about $33\times33$ kpc at the mean redshift $z=4.7$ of the survey, centered on the position of the source in the UV rest-frame  based on HST-814W images.  The object name and morpho-kinematic Class (see Sect.~\ref{class}) are indicated on top of each sub-panel.}
         \label{Fig_class1}
   \end{figure*}

\begin{figure*}
   \centering
  \includegraphics[width=\hsize]{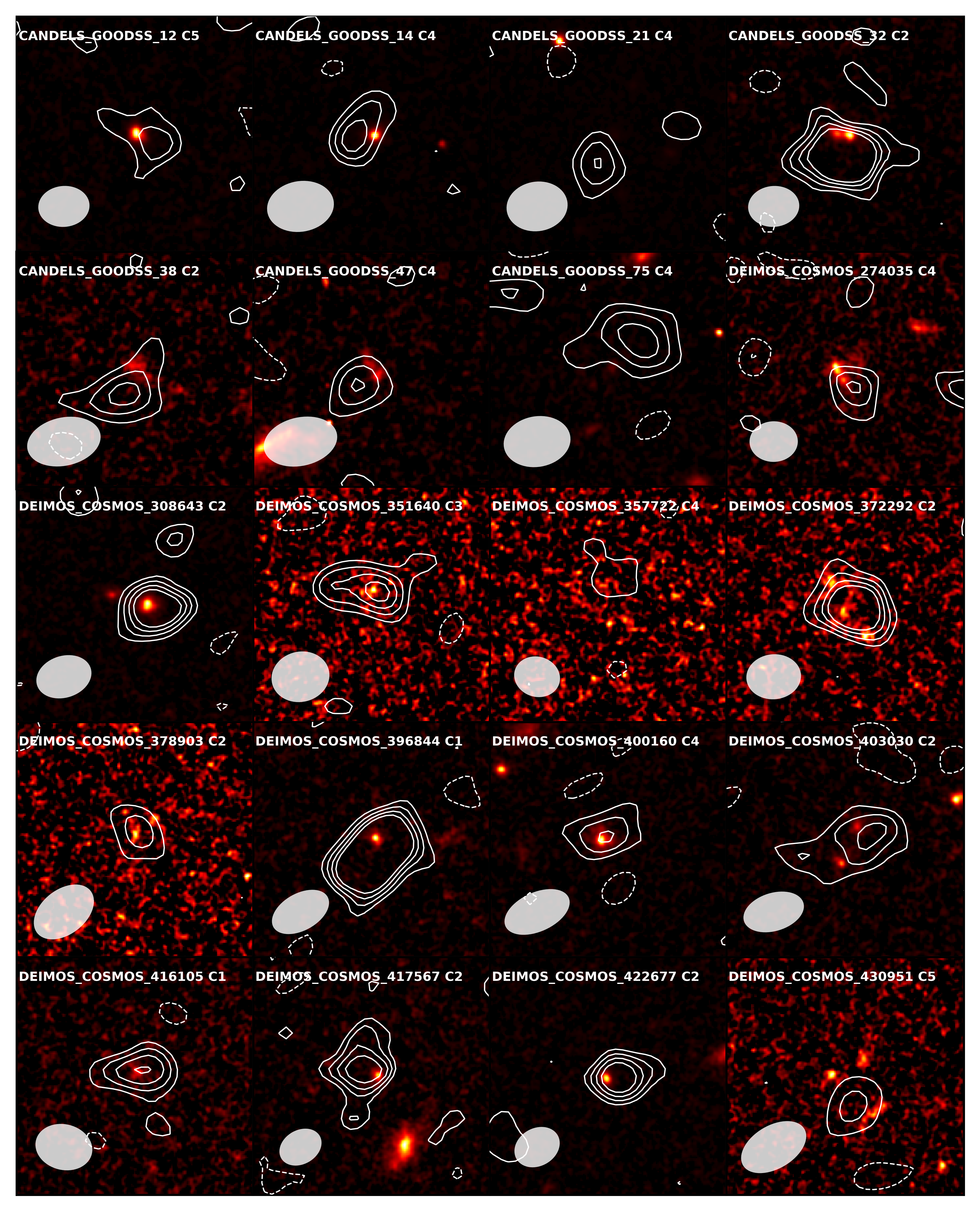}
      \caption{HST F814W images \citep{Koekemoer:07, Koekemoer:11} corresponding to [CII] flux maps in Fig.~\ref{Fig_class1}. Each panel is $5\arcsec \times 5\arcsec$ or about $33\times33$ kpc at the mean redshift $z=4.7$ of the survey, centered on the position of the source in the UV rest-frame HST-814W images. The white contours represent the distribution of the [CII] flux from Fig.~\ref{Fig_class1}, and the grey-filed ellipse is the ALMA beam size.}
         \label{Fig_hst1}
   \end{figure*}

\begin{figure*}
   \centering
  \includegraphics[width=\hsize]{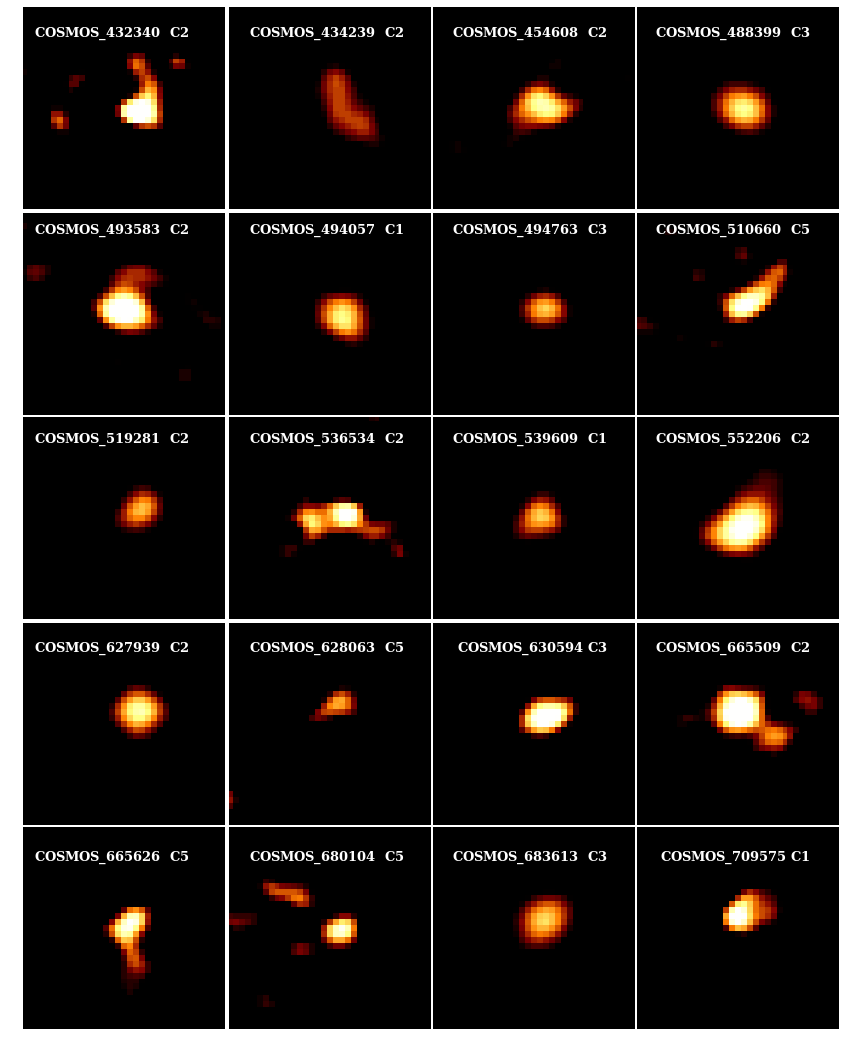}
      \caption{Velocity-integrated [C II] flux maps obtained collapsing the cube channels containing the [C II] line (see text). Each panel is  $5\arcsec \times 5\arcsec$ or about $33\times33$ kpc at the mean redshift $z=4.7$ of the survey, and centered on the position of the source in the UV rest-frame  based on HST-814W images.   The object name and morpho-kinematic Class (see Sect.~\ref{class}) are indicated on top of each sub-panel.}
         \label{Fig_class2}
   \end{figure*}

\begin{figure*}
   \centering
  \includegraphics[width=\hsize]{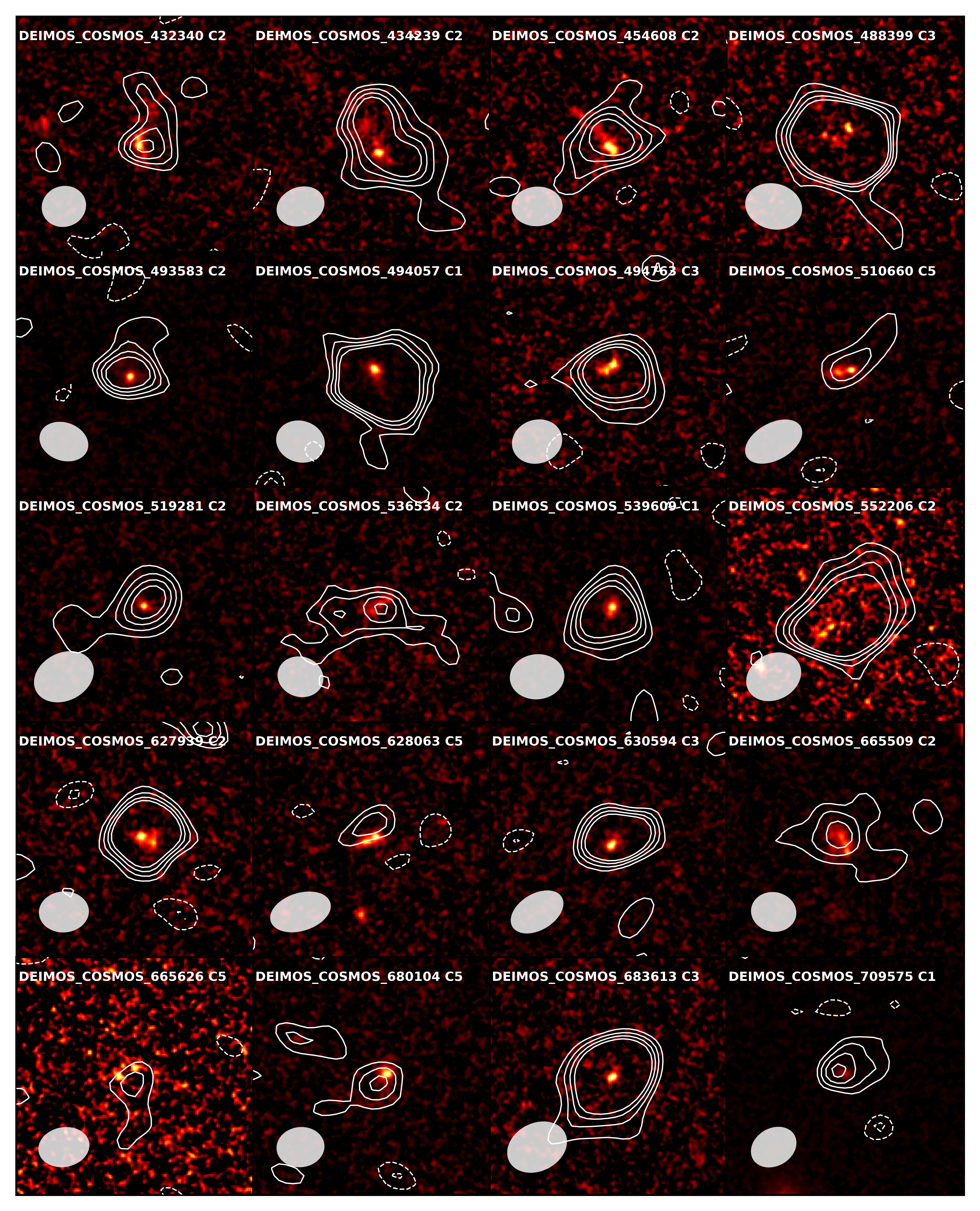}
      \caption{HST F814W images \citep{Koekemoer:07, Koekemoer:11} corresponding to [CII] flux maps in Fig.~\ref{Fig_class2}. Each panel is $5\arcsec \times 5\arcsec$ or about $33\times33$ kpc at the mean redshift $z=4.7$ of the survey, centered on the position of the source in the UV rest-frame HST-814W images. The white contours represent the distribution of the [CII] flux from Fig.~\ref{Fig_class2}, and the grey-filed ellipse is the ALMA beam size.}
         \label{Fig_hst2}
   \end{figure*}

\begin{figure*}
   \centering
  \includegraphics[width=\hsize]{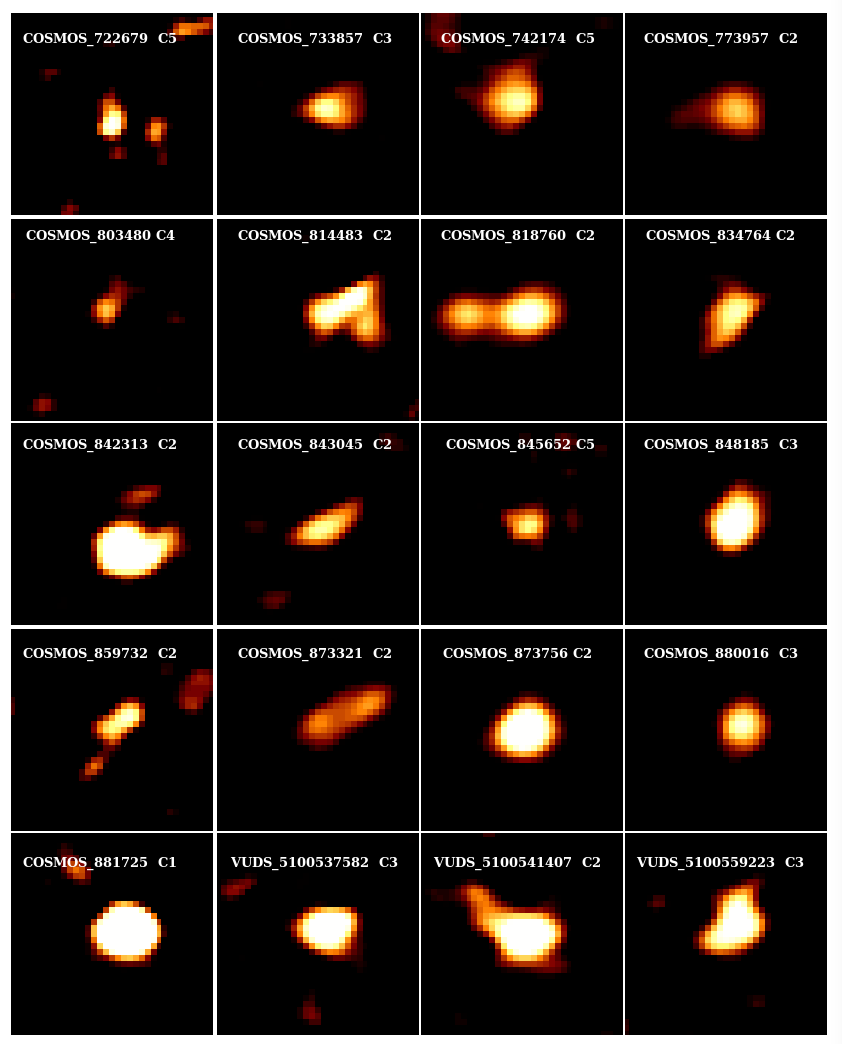}
      \caption{Velocity-integrated [C II] flux maps obtained collapsing the cube channels containing the [C II] line (see text). Each panel is  $5\arcsec \times 5\arcsec$ or about $33\times33$ kpc at the mean redshift $z=4.7$ of the survey, and centered on the position of the source in the UV rest-frame  based on HST-814W images. The object name and morpho-kinematic Class (see Sect.~\ref{class}) are indicated on top of each sub-panel.
              }
         \label{Fig_class3}
   \end{figure*}

\begin{figure*}
   \centering
  \includegraphics[width=\hsize]{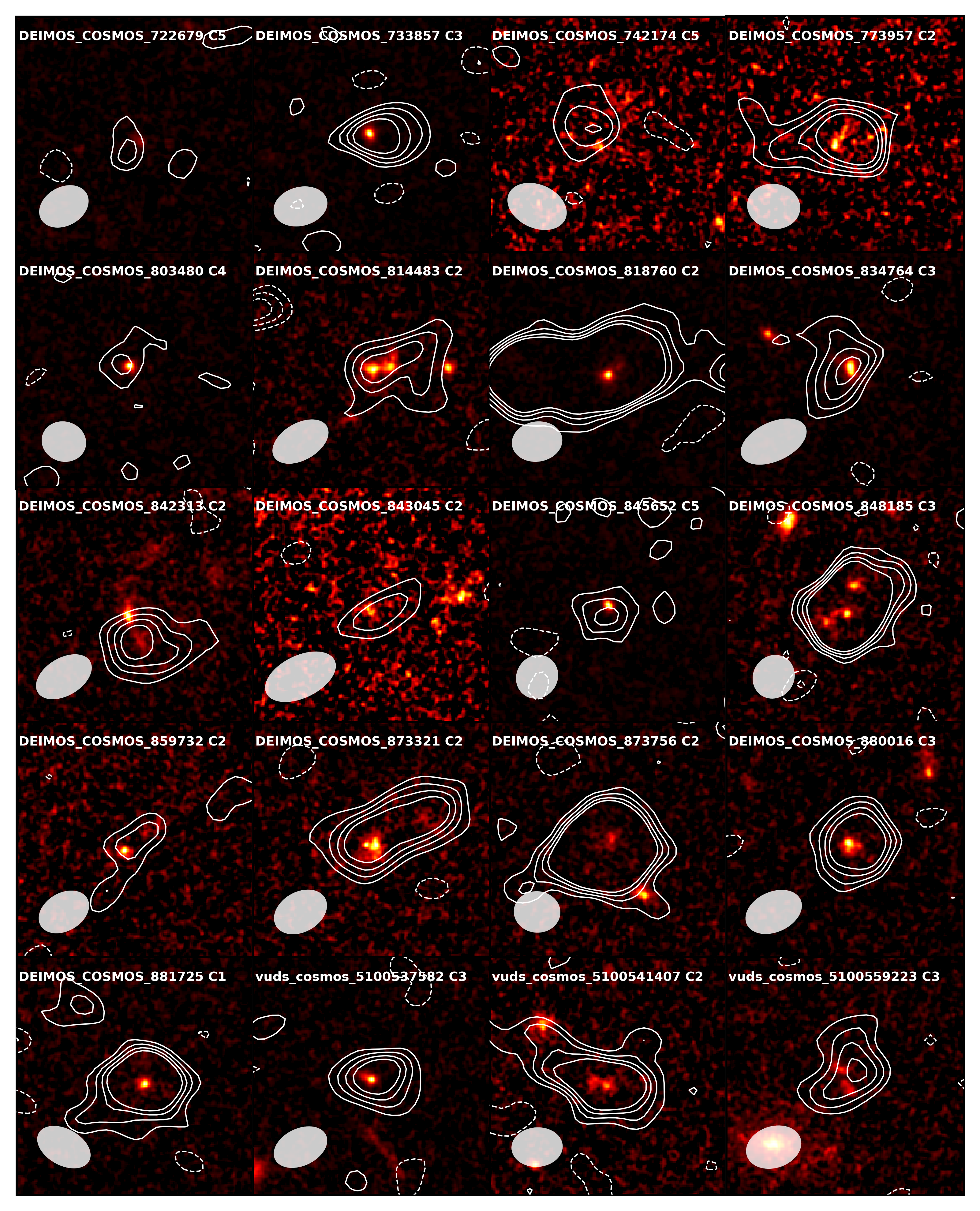}
      \caption{HST F814W images \citep{Koekemoer:07, Koekemoer:11} corresponding to [CII] flux maps in Fig.~\ref{Fig_class3}. Each panel is $5\arcsec \times 5\arcsec$ or about $33\times33$ kpc at the mean redshift $z=4.7$ of the survey, centered on the position of the source in the UV rest-frame HST-814W images. The white contours represent the distribution of the [CII] flux from Fig.~\ref{Fig_class3}, and the grey-filed ellipse is the ALMA beam size.}
         \label{Fig_hst3}
   \end{figure*}

\begin{figure*}
   \centering
  \includegraphics[width=\hsize]{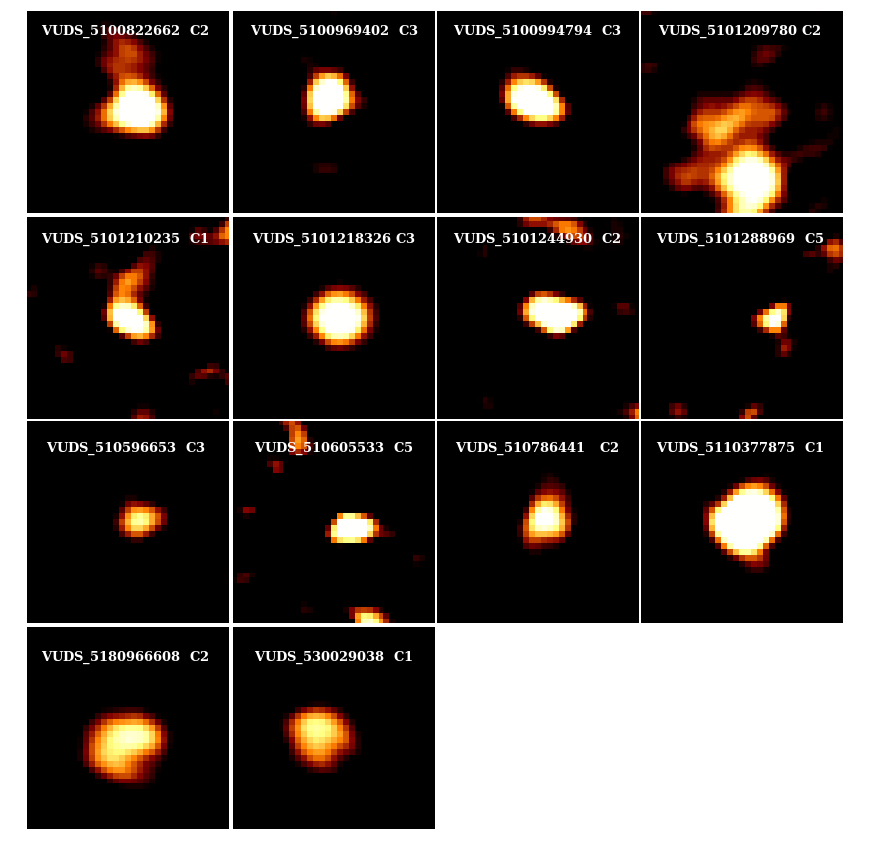}
      \caption{Velocity-integrated [C II] flux maps obtained collapsing the cube channels containing the [C II] line (see text). Each panel is  $5\arcsec \times 5\arcsec$ or about $33\times33$ kpc at the mean redshift $z=4.7$ of the survey, and centered on the position of the source in the UV rest-frame  based on HST-814W images.    The object name and morpho-kinematic Class (see Sect.~\ref{class}) are indicated on top of each sub-panel.}
         \label{Fig_class4}
   \end{figure*}

\begin{figure*}
   \centering
  \includegraphics[width=\hsize]{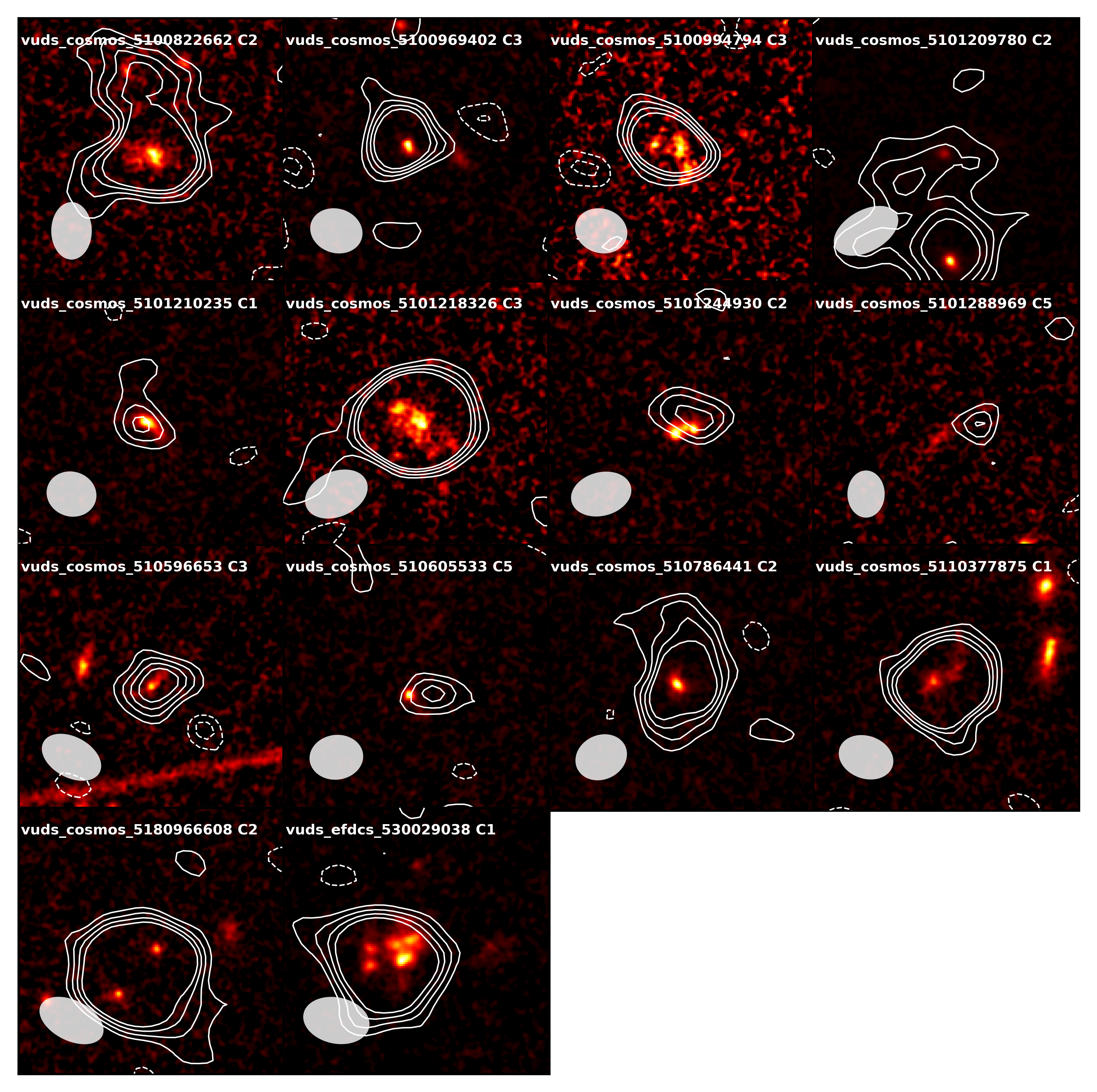}
      \caption{HST  F814W images \citep{Koekemoer:07, Koekemoer:11} corresponding to [CII] flux maps in Fig.~\ref{Fig_class4}. Each panel is $5\arcsec \times 5\arcsec$ or about $33\times33$ kpc at the mean redshift $z=4.7$ of the survey, centered on the position of the source in the UV rest-frame HST-814W images. The white contours represent the distribution of the [CII] flux from Fig.~\ref{Fig_class4}, and the grey-filed ellipse is the ALMA beam size.}
         \label{Fig_hst4}
   \end{figure*}

Images in the [CII] line of sources with [CII] detected at more than $3.5\sigma$ are presented in Fig.~\ref{Fig_class1}, \ref{Fig_class2}, \ref{Fig_class3},   and Fig.~\ref{Fig_class4}. These so-called [CII] "flux images" (flux maps hereafter) are produced using the CLEAN algorithm and the ~{\texttt immoments} routine, optimally extracted using an iterative process to determine the line profile of the source and then collapse channels [$f_{cen} - FWHM$; $f_{cen} + FWHM$], where $f_{cen}$ is the central frequency of [CII], which can be a posteriori confirmed to be different from the UV (see \citealt{Bethermin:20}).   These flux maps are compared to 
HST i-band F814W images \citep{Koekemoer:07, Koekemoer:11} in Figs .~\ref{Fig_hst1}, \ref{Fig_hst2}, \ref{Fig_hst3}, and \ref{Fig_hst4}), representing the UV rest-frame, with [CII] contours overlaid. 

These images give a first view of the shape of the [CII] emission in normal galaxies at 4<z<6. There are several facts worth noting. Even though the observations were carried out with a beam size providing moderate spatial resolution with FWHM$\sim$0.7\arcsec, about two thirds of the sources are resolved in [CII]. This means that intrinsic (total) sizes as seen in atomic gas must be about the size of the beam, or a significant fraction thereof, hence physical source sizes reaching several kilo-parsecs. By itself this fact gives an indication that physical processes at work in those galaxies are puffing up their sizes beyond being compact. Another striking evidence from these images is the large diversity of   [CII] emission morphology. Some objects appear as very extended (e.g., bottom-left object COSMOS 881725 in Fig.~\ref{Fig_class3}), some others with double merger-like components (e.g., third row, second-from-left object COSMOS 351640 in Fig.~\ref{Fig_class1}), while others are compact (unresolved). This diversity must also reflect a diversity in the physical processes at work.

We focus more on this in attempting a visual classification in the following section, Sect.~\ref{class}, and future papers will concentrate on the quantitative properties of the [CII] emission in the different classes of the ALPINE sample.

%----------------------------------------------------------------- 

%----------------------------------------------------------------- 
%-----------------------------------------------------------------

%----------------------------------------------------------------- 
%-----------------------------------------------------------------
%----------------------------------------------------------------- 
%--------------------------------------------------------------------
%----------------------------------------------------------------- 
%--------------------------------------------------------------------
\section{Morpho-Kinematic  classification}
\label{class}

As presented in Sect.~\ref{c2maps}, the [CII] emission appears spatially very diverse. As we have at our disposal not only the flux  maps, but also the full ($\alpha, \delta$, velocity) 3D data-cubes for all sources as well as all ancillary information presented in Sect.~\ref{ancillary}, we are able to perform an empirical visual-based morpho-kinematic classification, as discussed below.

\begin{figure*}
   \centering
  \includegraphics[width=\hsize]{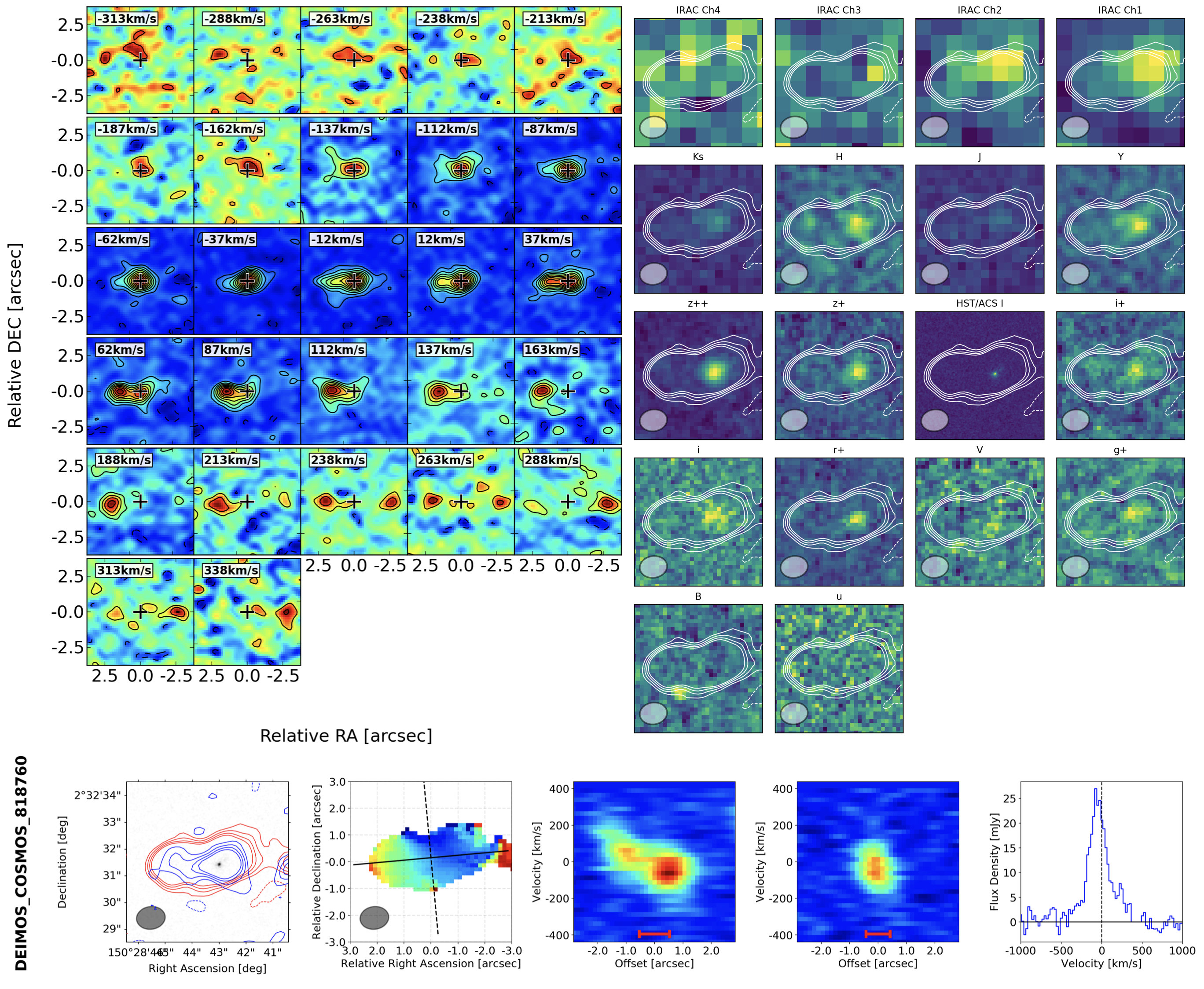}
      \caption{Summary panel with all information available for the classification of galaxies. {\it Top left:} Velocity channel maps, in 25km/s velocity intervals. {\it Top right:} Optical and NIR images, with contours from the [CII] emission. {\it Bottom, left to right:} Flux  map in [CII] (red contours) and continuum emission (blue contours) overlaid on top of the i-band F814W HST image; velocity map with major and minor axes used to produce the PV diagram of the next two panels on the right; [CII] line emission in velocity (V=0 being from the UV-derived spectroscopic redshift). This galaxy has been classified as a merger (Class 2).
              }
         \label{Fig_data}
   \end{figure*}

The observational data were assembled in one slide per object, including the ALMA data  with flux  map, the channel maps cut in 25 km/s velocity intervals, the velocity field map (moment 1), position-velocity (PV) diagrams projected along the major and minor axes of the velocity map, and the integrated 1D [CII] spectrum, together with multi-band optical and NIR images including HST ACS F814W i-band and/or WFC3-F160W images when available, as presented in Fig.~\ref{Fig_data}. All these data served as input to object classification described below.

Following previous work from 3D integral field spectroscopy using H$\alpha$ \citep[e.g.,][]{Forster2009,Epinat2012},  we define the following morpho-kinematic classes:
\begin{itemize}
\item Class 1: Rotator. This class is defined for galaxies satisfying the following criteria:
\begin{itemize}
\item Smooth transition between intensity channel maps
\item Obvious gradient in velocity field map (Moment 1)
\item Tilted  PV along the major axis, straight  PV on the minor axis
\item Possible double-horned profile in spectrum
\item Single component in ancillary data
\end{itemize}
\item Class 2: Pair-Merger (major or minor), interacting system:
\begin{itemize}
\item Complex behavior in channel maps
\item Separate components in flux  maps or PV diagrams
\item Multiple components in ancillary data
\end{itemize}
\item Class 3: Extended Dispersion dominated
\begin{itemize}
\item No positional shift of emission across intensity channel maps
\item Straight PV diagrams
\item Extended beyond the ALMA beam in flux  maps
\end{itemize}
\item Class 4: Compact Dispersion dominated
\begin{itemize}
\item No positional shift of emission across intensity channel maps
\item Straight PV diagrams
\item Unresolved in flux maps
\end{itemize}
\item Class 5: too weak to be classified
\end{itemize}

All the observational material described above was then visually inspected, independently, by eight people in the team, each assigning a class to each ALPINE object. This provided a statistical basis to estimate the mode of the classification for each object and a rough dispersion obtained as the average of the difference between the mode and each individually-measured class. To mitigate somewhat  the well-known effect of a dominant class not necessarily being "the truth" (e.g., most people missing some key evidence while only a few spotted it), the mode and extremes of the classification were examined by two people who proposed to the team the  class satisfying all identified evidence. One last iteration was then performed, with individual team members asked to identify the objects for which they were in most disagreement with, followed by a last round to agree on a final classification. This led to the final classification listed in Table ~\ref{tab_prop}.

In order to classify these galaxies using our qualitative methods, bright line emission must be present in multiple consecutive channels. For a given S/N, a narrow line will thus be much easier to classify than a broad line, as the S/N per channel is higher. This is seen in VUDS\_COSMOS\_510605533 (S/N$_{\rm [CII]}$=4.9, class 5, broad line) vs CANDELS\_GOODSS\_21 (/SN$_{\rm [CII]}$=4.2, class 4, narrow line). While the S/N$_{\rm CII}$ in Table~\ref{tab_prop} is an indicator of the overall line strength, the classification depends on both the S/N$_{\rm [CII]}$ and FWHM.

It is also possible that noise peaks in the data cube could be interpreted as galaxy components, earning the galaxy a merger classification (class 2). To reduce this error, we examined the integrated spectrum, moment maps, and channel maps, ignoring emission peaks that were narrow and kinematically distant from the galaxy. It should be noted that simulations of z$>$4 galaxies \citep[e.g.,][]{Pallottini2017,Kohandel2019} show many minor satellites around each galaxy on small scales, which suggests that our relatively low resolution survey may underestimate the true number of mergers.

\begin{figure}
   \centering
  \includegraphics[width=\hsize]{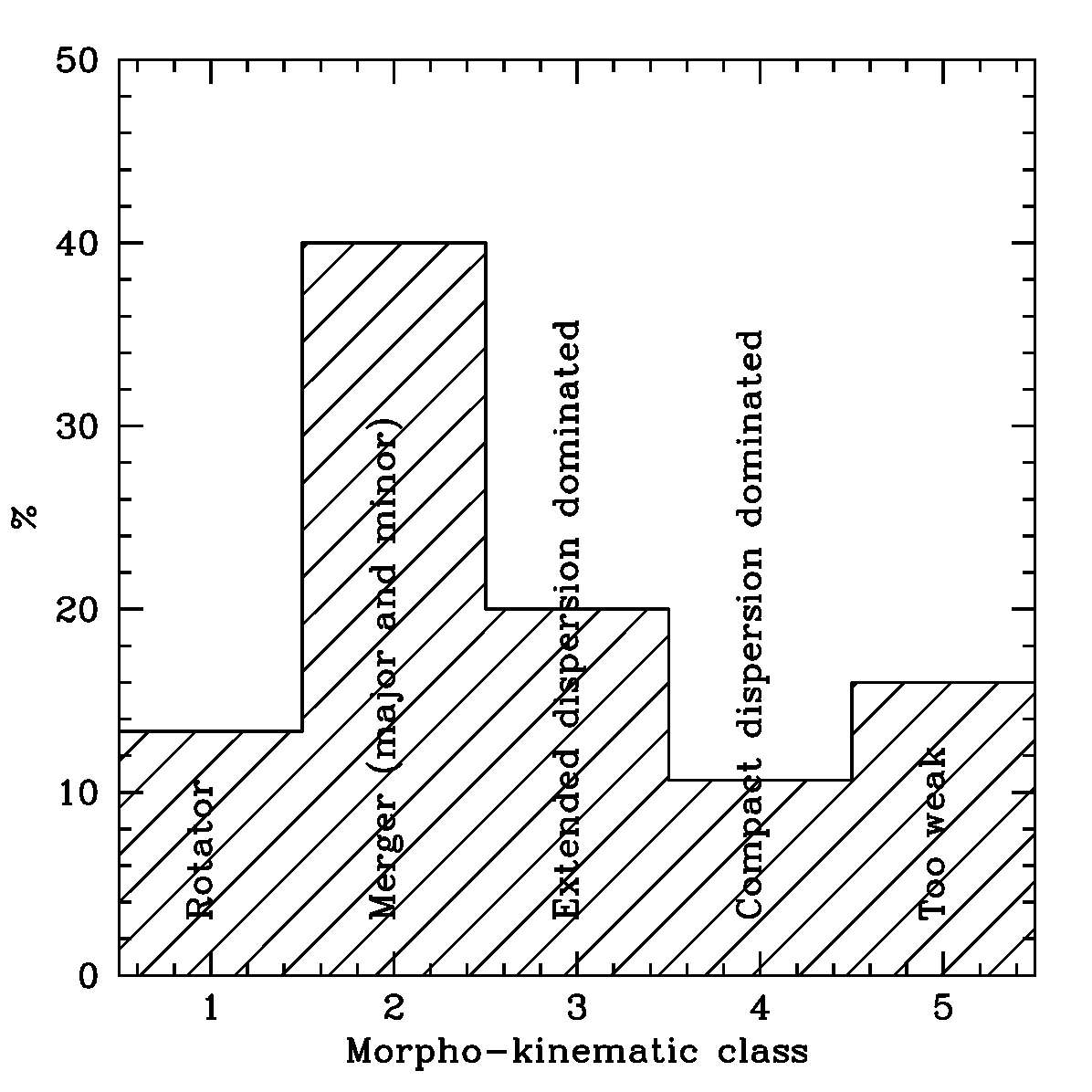}
      \caption{Distribution of morpho-kinematic classes in the ALPINE sample for sources with [CII] measured at more than 3.5$\sigma$ above the noise.
              }
         \label{Fig_class}
   \end{figure}

The distribution in the different classes is presented in Fig.\ref{Fig_class}. We find 13.3\% rotating discs in class 1, 40\% of galaxies in the merger class 2, 20\% in the extended and dispersion dominated class 3, 10.7\% compact in class 4, and the remaining 16\% of the sample being too difficult to classify (class 5). If we consider only the S/N$_{\rm [CII]}>$5 objects, the distribution is slightly different: 17\,\% of rotating discs, 51\,\% of mergers, 32\,\% of extended and dispersion-dominated systems, 0\,\% of compact systems, and 0\,\% of objects too faint to be classified. It is not surprising that there is no object in this last category at higher S/N. No object is also found in the compact class. This may be caused by the higher S/N threshold, which could be biased against lower mass and lower SFR systems. We note that the relative contribution of the three other classes (1, 2, and 3) does not change significantly ($<$2\,$\sigma$). We find that while our sample contains both mergers and rotators, there is also a large number of dispersion-dominated sources. This diversity of kinematic classes for a sample of galaxies with similar SFRs and UV characteristics suggests that the evolutionary tracks of these galaxies in the early Universe (0.9-1.4\,Gyr after the Big Bang) already had significant variations. Future systematic studies of the morpho-kinematic classification as a function of the galaxy properties will be presented in future papers (Romano et al. in prep., Jones et al. in prep.). 

We note the high fraction of mergers (40\,\%), which indicates that mass assembly through merging is frequent at these redshifts for normal main sequence SFGs. This value is significantly higher than that of \citet[$8.3_{-3.7}^{+7.6}\,,\%$ at 4$<$z$<$6]{ventou2017}, but their objects have log(M$_\star$)$<9.5$ while most of ALPINE objects are more massive. The methods for classifying mergers and the observed wavelength are also very different. A preliminary examination of spatial and velocity information indicates that most merging systems would merge within 0.5 to 1 Gyr \citep[e.g., the triple merger system presented by ][]{Jones2019}, which then means that most of these mergers would end up forming one single galaxy by $z\sim2.5$. Merging systems observed at sub-mm wavelengths have been reported previously, but for more starburst-like objects \citep[e.g., recently, ][]{Danielson2017,Tadaki2018,Zhang2018,diaz2018,Hodge2019}. The presence of extended [CII] nebulae in class 3 is also quite striking as an indication  that large extended gas reservoirs were readily available to fuel star formation right after reionization ended.

Simulations and extrapolations from local studies suggest that the merger rate was high at z$>$4 \citep[e.g.,][]{Mantha2018,Kohandel2019} and targeted [CII] observations of normal z$>$4 galaxies have revealed some galaxies with clumpy morphologies \citep{Carniani2018} which are likely caused by mergers \citep{Calabro2019}. But several observations have revealed evidence for ordered rotation \citep[e.g.,][]{De_Breuck2014,Jones2017,Smit2018,Bakx2020,Tadaki2020}, so the true distribution of kinematic states has been unknown. For the first time, we can examine the kinematics of a statistically significant number of galaxies, in order to determine this.

The properties of galaxies in these different samples will be extensively described in future papers from the ALPINE team. It may well be that some objects in class 4 would appear as rotators when observed under higher spatial resolution, as seen in near-IR integrated field observations of $z\sim2$ massive galaxies \citep[e.g.,][]{Forster2018}. On the other hand, galaxies at z=1-3, with stellar masses comparable to the ALPINE sample, show a larger proportion of dispersion-dominated galaxies than galaxies with higher masses.
%----------------------------------------------------------------- 

%-----------------------------------------------------------------

%----------------------------------------------------------------- 
%-----------------------------------------------------------------

%--------------------------------------------------------------------
\section{Summary and conclusions}
\label{summary}

The ALMA-ALPINE [CII] survey (A2C2S) provides an unprecedented view of a representative sample of 118 star-forming galaxies observed in their assembly right after the end of HI reionization at redshifts 4<z<6. Galaxies are selected on the basis of an existing reliable spectroscopic redshift and using the SED-based SFR to predict the [CII] flux using the  \cite{DeLooze2014} relation and selecting SFR such that $L[CII] > 1.2\times10^8 L_{\odot}$. The overall detection rate is 64\% for galaxies detected in [CII] $3.5\sigma$ above the noise, 21\% in the continuum. We present the ALPINE survey strategy and sample properties, along with the the projected [CII] flux maps. Combining these maps with velocity channel maps, velocity field, and all available ancillary information, we  have established a classification scheme. We find a surprisingly wide range of galaxy types,
including 40\% mergers, 20\% extended and dispersion-dominated, 13.3\% rotating discs, and 10.7\% compact, with the remaining 16\% too faint to be classified. This diversity of types indicates that several physical processes are at work for the assembly of mass in these galaxies, first and foremost, for galaxy merging. While galaxy merging is commonly associated with starbursts above the main sequence, at least up to $z\sim3$, merging systems in ALPINE at $z\sim4.7$ lie mainly on the MS, and therefore, merging is also a dominant process for normal SFGs at this epoch. This will be further investigated in future papers.

The ALPINE sample offers a unique opportunity to study galaxies in the process of their  assembling. This paper is the first in a series and future papers will present analyses of specific populations, as well as general statistical properties. ALPINE galaxies are ideally suited for follow-ups with such as the James Webb Space Telescope (JWST) and   the Extremely
Large Telescopes (ELTs).

\begin{acknowledgements}
This paper is based on data obtained with the ALMA Observatory, under the Large Program 2017.1.00428.L. ALMA is a partnership of ESO (representing its member states), NSF(USA) and NINS (Japan), together with NRC (Canada), MOST and ASIAA (Taiwan), and KASI (Republic of Korea), in cooperation with the Republic of Chile. The Joint ALMA Observatory is operated by ESO, AUI/NRAO and NAOJ. This program receives funding from the CNRS national program Cosmology and Galaxies. AC, FP, MT, CG acknowledge the support from grant PRIN MIUR 2017. G.C.J. and R.M. acknowledge ERC Advanced Grant 695671 ``QUENCH'' and support by the Science and Technology Facilities Council (STFC). E.I.\ acknowledges partial support from FONDECYT through grant N$^\circ$\,1171710. The Cosmic Dawn Center (DAWN) is funded by the Danish National Research Foundation under grant No. 140 . S.T. acknowledges support from the ERC Consolidator Grant funding scheme (project Context, grant No. 648179). LV acknowledges funding from the European Union’s Horizon 2020 research and innovation program under the Marie Sklodowska-Curie Grant agreement No. 746119. D.R. acknowledges support from the National Science Foundation under grant numbers AST-1614213 and AST-1910107 and from the Alexander von Humboldt Foundation through a Humboldt Research Fellowship for Experienced Researchers. JDS was supported by the JSPS KAKENHI Grant Number JP18H04346, and the World Premier International Research Center Initiative (WPI Initiative), MEXT, Japan. GL acknowledges support from the European Research Council (ERC) under the European Union’s Horizon 2020 research and innovation programme (project CONCERTO, grant agreement No 788212) and from the Excellence Initiative of Aix-Marseille University-A*Midex, a French “Investissements d’Avenir” programme.
\end{acknowledgements}


\bibliographystyle{aa}
\bibliography{refs_alpine}

\clearpage
\onecolumn

\begin{longtable}{l c c c c c}
%\begin{table*}[h]
\caption{\label{tab_prop} Properties of the ALPINE sample of galaxies with [CII] detected above 3.5$\sigma$, including position, spectroscopic redshift $z_{spec}$ derived from the UV spectra, S/N, as well as the object class as described in Sect.~6}\\
\hline
            \noalign{\smallskip}
            ALPINE ID & $\alpha_{2000}$ & $\delta_{2000}$ & $z_{spec}$  & S/N([CII]) &  morpho-kinematic class  \\ \hline
            \noalign{\smallskip}
\endfirsthead

\caption{continued.}\\
\hline
           \noalign{\smallskip}
            ALPINE ID & $\alpha_{2000}$ & $\delta_{2000}$ & $z_{spec}$  & S/N([CII]) &  morpho-kinematic class  \\ \hline
            \noalign{\smallskip}
\endhead
\hline
\endfoot
\centering             
%Object ID & RA & DEC & z_{spec} & SNR [CII] & morpho-kinematic class \\ \hline
CANDELS\_GOODSS\_12 & 53.2251 & -27.8336 & 4.4297 & 4.4 & 5\\ 
CANDELS\_GOODSS\_14 & 53.0788 & -27.8841 & 5.5630 & 4.6 & 4\\ 
CANDELS\_GOODSS\_21 & 53.0497 & -27.6992 & 5.5780 & 4.2 & 4\\ 
CANDELS\_GOODSS\_32 & 53.0708 & -27.6871 & 4.4140 & 12.3 & 2\\ 
CANDELS\_GOODSS\_38 & 53.0662 & -27.6900 & 5.5740 & 4.7 & 2\\ 
CANDELS\_GOODSS\_42 & 53.1659 & -27.8828 & 5.5430 & 3.7 & 5\\ 
CANDELS\_GOODSS\_47 & 53.1885 & -27.8194 & 5.5830 & 4.0 & 4\\ 
CANDELS\_GOODSS\_75 & 53.1359 & -27.7983 & 5.6000 & 4.8 & 4\\ 
DEIMOS\_COSMOS\_274035 & 149.8853 & 1.7017 & 4.4820 & 4.4 & 4\\ 
DEIMOS\_COSMOS\_308643 & 150.3612 & 1.7573 & 4.5270 & 7.7 & 2\\ 
DEIMOS\_COSMOS\_351640 & 150.3712 & 1.8248 & 5.7070 & 5.7 & 3\\ 
DEIMOS\_COSMOS\_357722 & 149.9668 & 1.8349 & 5.7380 & 3.6 & 4\\ 
DEIMOS\_COSMOS\_372292 & 149.9132 & 1.8578 & 5.1370 & 9.6 & 2\\ 
DEIMOS\_COSMOS\_378903 & 150.2976 & 1.8684 & 5.4300 & 4.6 & 2\\ 
DEIMOS\_COSMOS\_396844 & 150.2485 & 1.8965 & 4.5400 & 12.1 & 1\\ 
DEIMOS\_COSMOS\_400160 & 150.2671 & 1.9014 & 4.5330 & 4.5 & 4\\ 
DEIMOS\_COSMOS\_403030 & 150.0274 & 1.9059 & 4.5680 & 5.0 & 2\\ 
DEIMOS\_COSMOS\_416105 & 150.6903 & 1.9266 & 5.6340 & 5.3 & 1\\ 
DEIMOS\_COSMOS\_417567 & 150.5171 & 1.9289 & 5.6750 & 6.4 & 2\\ 
DEIMOS\_COSMOS\_422677 & 150.4978 & 1.9369 & 4.4430 & 7.1 & 2\\ 
DEIMOS\_COSMOS\_430951 & 150.3268 & 1.9510 & 5.6840 & 4.1 & 5\\ 
DEIMOS\_COSMOS\_432340 & 150.5398 & 1.9516 & 4.4070 & 5.5 & 2\\ 
DEIMOS\_COSMOS\_434239 & 150.3213 & 1.9553 & 4.4900 & 7.4 & 2\\ 
DEIMOS\_COSMOS\_454608 & 150.6807 & 1.9891 & 4.5780 & 6.5 & 2\\ 
DEIMOS\_COSMOS\_488399 & 150.7548 & 2.0433 & 5.6780 & 26.2 & 3\\ 
DEIMOS\_COSMOS\_493583 & 150.0973 & 2.0512 & 4.5160 & 8.3 & 2\\ 
DEIMOS\_COSMOS\_494057 & 149.6188 & 2.0519 & 5.5400 & 17.1 & 1\\ 
DEIMOS\_COSMOS\_494763 & 150.0213 & 2.0534 & 5.2380 & 10.5 & 3\\ 
DEIMOS\_COSMOS\_510660 & 149.9715 & 2.0771 & 4.5540 & 4.0 & 5\\ 
DEIMOS\_COSMOS\_519281 & 149.7538 & 2.0910 & 5.5700 & 6.7 & 2\\ 
DEIMOS\_COSMOS\_536534 & 149.9719 & 2.1182 & 5.6940 & 5.0 & 2\\ 
DEIMOS\_COSMOS\_539609 & 149.7803 & 2.1226 & 5.1700 & 8.9 & 1\\ 
DEIMOS\_COSMOS\_552206 & 149.6116 & 2.1409 & 5.5140 & 14.8 & 2\\ 
DEIMOS\_COSMOS\_627939 & 150.2703 & 2.2539 & 4.5320 & 13.0 & 2\\ 
DEIMOS\_COSMOS\_628063 & 150.2177 & 2.2543 & 4.5390 & 3.8 & 5\\ 
DEIMOS\_COSMOS\_630594 & 150.1359 & 2.2579 & 4.4470 & 11.2 & 3\\ 
DEIMOS\_COSMOS\_665509 & 149.7352 & 2.3109 & 4.5290 & 4.8 & 2\\ 
DEIMOS\_COSMOS\_665626 & 150.3093 & 2.3118 & 4.5830 & 4.4 & 5\\ 
DEIMOS\_COSMOS\_680104 & 150.2923 & 2.3323 & 4.5320 & 4.2 & 5\\ 
DEIMOS\_COSMOS\_683613 & 150.0393 & 2.3372 & 5.5360 & 13.6 & 3\\ 
DEIMOS\_COSMOS\_709575 & 149.9461 & 2.3758 & 4.4150 & 5.5 & 1\\ 
DEIMOS\_COSMOS\_722679 & 149.9371 & 2.3962 & 5.7580 & 4.0 & 5\\ 
DEIMOS\_COSMOS\_733857 & 150.3330 & 2.4132 & 4.5470 & 7.3 & 3\\ 
DEIMOS\_COSMOS\_742174 & 150.1630 & 2.4257 & 5.6410 & 4.8 & 5\\ 
DEIMOS\_COSMOS\_773957 & 150.2919 & 2.4747 & 5.6830 & 8.5 & 2\\ 
DEIMOS\_COSMOS\_803480 & 149.9885 & 2.5202 & 4.5420 & 3.7 & 4\\ 
DEIMOS\_COSMOS\_814483 & 150.3632 & 2.5362 & 4.5840 & 4.6 & 2\\ 
DEIMOS\_COSMOS\_818760 & 150.4786 & 2.5421 & 4.5540 & 26.7 & 2\\ 
DEIMOS\_COSMOS\_834764 & 149.8989 & 2.5668 & 4.5040 & 5.4 & 3\\ 
DEIMOS\_COSMOS\_842313 & 150.2272 & 2.5764 & 4.5520 & 6.5 & 2\\ 
DEIMOS\_COSMOS\_843045 & 150.0515 & 2.5788 & 5.8170 & 4.1 & 2\\ 
DEIMOS\_COSMOS\_845652 & 150.2150 & 2.5827 & 5.3100 & 4.9 & 5\\ 
DEIMOS\_COSMOS\_848185 & 150.0896 & 2.5864 & 5.2840 & 18.3 & 3\\ 
DEIMOS\_COSMOS\_859732 & 150.0021 & 2.6053 & 4.5340 & 4.3 & 2\\ 
DEIMOS\_COSMOS\_873321 & 150.0169 & 2.6266 & 5.1580 & 7.5 & 2\\ 
DEIMOS\_COSMOS\_873756 & 150.0113 & 2.6278 & 4.5480 & 32.6 & 2\\ 
DEIMOS\_COSMOS\_880016 & 149.9799 & 2.6356 & 4.5380 & 8.6 & 3\\ 
DEIMOS\_COSMOS\_881725 & 150.0565 & 2.6380 & 4.5850 & 12.3 & 1\\ 
vuds\_cosmos\_5100537582 & 150.3897 & 1.8390 & 4.5460 & 8.1 & 3\\ 
vuds\_cosmos\_5100541407 & 150.2538 & 1.8094 & 4.5485 & 11.4 & 2\\ 
vuds\_cosmos\_5100559223 & 150.2214 & 1.8648 & 4.5577 & 5.9 & 3\\ 
vuds\_cosmos\_5100822662 & 149.7413 & 2.0810 & 4.5235 & 14.9 & 2\\ 
vuds\_cosmos\_5100969402 & 150.3338 & 2.2836 & 4.5869 & 11.0 & 3\\ 
vuds\_cosmos\_5100994794 & 150.1715 & 2.2873 & 4.5783 & 12.0 & 3\\ 
vuds\_cosmos\_5101209780 & 150.3894 & 2.3695 & 4.5700 & 4.3 & 2\\ 
vuds\_cosmos\_5101210235 & 150.3817 & 2.3661 & 4.5733 & 4.3 & 1\\ 
vuds\_cosmos\_5101218326 & 150.3021 & 2.3146 & 4.5678 & 26.6 & 3\\ 
vuds\_cosmos\_5101244930 & 150.1986 & 2.3006 & 4.5769 & 5.0 & 2\\ 
vuds\_cosmos\_5101288969 & 149.8777 & 2.3316 & 5.6971 & 4.2 & 5\\ 
vuds\_cosmos\_510596653 & 149.8262 & 1.9381 & 4.5655 & 6.2 & 3\\ 
vuds\_cosmos\_510605533 & 149.8526 & 1.8786 & 4.5065 & 4.9 & 5\\ 
vuds\_cosmos\_510786441 & 150.1429 & 1.9892 & 4.4618 & 11.1 & 2\\ 
vuds\_cosmos\_5110377875 & 150.3847 & 2.4084 & 4.5441 & 18.5 & 1\\ 
vuds\_cosmos\_5180966608 & 150.4061 & 2.1399 & 4.5293 & 12.5 & 2\\ 
vuds\_efdcs\_530029038 & 53.0792 & -27.8772 & 4.4179 & 9.2 & 1\\ 
\noalign{\smallskip}
% \end{tabular}
 \end{longtable}

\end{document}